\documentclass[11pt,twoside,a4paper]{article}
\usepackage{mathtools}
\usepackage{amsmath,amssymb,amsthm}
\usepackage{hyperref}
\usepackage{graphicx,psfrag}
\usepackage{nicefrac}
\usepackage{tikz}
\usepackage{enumitem}

\newtheorem{thm}{Theorem}[section]
\newtheorem{dfn}[thm]{Definition}
\newtheorem{tad}[thm]{Theorem and Definition}

\theoremstyle{definition}

\theoremstyle{remark}
\newtheorem{rem}[thm]{Remark}

\newcommand{\R}{\mathbb{R}}
\newcommand{\Z}{\mathbb{Z}}

\renewcommand{\phi}{\varphi}

\newcommand{\charact}{\operatorname{char}}
\newcommand{\determ}{\operatorname{det}}
\newcommand{\dr}{\operatorname{Dr}}

\begin{document}

\title{Design of Ciphers based on the Geometric Structure of the Laguerre and Minkowski Planes}
\author{Christoph Capellaro\footnote{EY EMEIA Financial Services - Cyber Security}}

\date{}
\maketitle

 \begin{abstract}
 	Till now geometric structures don't play a major role in cryptography. Gilbert, MacWilliams and Sloane~\cite{GMS74} introduced an authentication scheme in the projective plane and showed its perfectness in the sense of Shannon~\cite{Sha49}. In \cite{Cap21} we introduced an encryption scheme in the M\"obius plane and showed that it fulfills Shannon's requirement of perfectness in first approximation and also the requirement of completeness according to Kam and Davida~\cite{KD79}. In this paper we will apply a similar approach to define encryption schemes in the geometries of the Laguerre plande and the Minkowski plane. We will show that the encryption scheme in the Laguerre geometry meets Shannon's requirement of perfectness sharp and that the encryption scheme in the Minkowski geometry meets this requirement in first approximation. The Laguerre cipher also fulfills the requirement of completeness according to Kam and Davida. 
 \end{abstract}


\medskip

\section{Introduction}	
 	A cryptographic transformation can be understood as an incidence relation, whereby messages ${\text m}$ and ciphertexts ${\text c}$ are represented as points. The cryptographic transformation $f$ that maps ${\text m}$ to ${\text c}$ is then described by a geometric object that incises with these two points. In this paper we will design new encryption transformations in the geometries of the Laguerre plane and the Minkowski plane and analyze their properties.
 	
The basic property of cycles in the Laguerre and Minkowski plane, being that three points incise with a cycle, allow to associate one point with a message, another point with the ciphertext and still have a degree of freedom for a secret key. We will analyze encryption methods in the Laguerre and Minkowski plane, namely the criteria of perfectness according to Shannon~\cite{Sha49} and the completeness in the case that the geometry is defined over a field of characteristic 2. Shannon's requirement of perfectness states that basically the result of the encryption process cannot be distinguished from a noisy channel. Completeness according to Kam and Davida~\cite{KD79} means that, assumed that input and output of a transformation are represented as bit vectors, there is at least one input vector for which a change in the $i$-th bit results in a change of the $j$-th bit of the output vector for arbitrary $i$ and $j$.  
 
 {\it Aknowledgment.} The author would like to express great gratitude to Helmut Karzel. Without the insights in circle geometries that he shared, this paper would not have been possible. Furthermore thank goes to Mariia Denysenko, who with her inspiring attitude and due care of technical details supported the completion of this paper.
 
\section{Cryptographic Transformations and Cryptographic Schemes} 
This and the following section repeat definitions and theorems from~\cite{Cap21}. They will be required to define cryptographic schemes in the Laguerre and Minkowski plane and to discuss their properties.
\begin{dfn}\label{dfn:cryptographic_transformation} Let $M$ and $C$ be sets,
	where $M$ is called a set of messages and $C$ a set of ciphertexts. If there is a nonempty set of functions $F$ of the form $F:M\to C$ with the property that every $f\in F$ is reversible, then the elements $f\in F$ are called {\em cryptographic transformations} and $(M,C,F)$ is called a {\em cryptographic scheme}.
\end{dfn}

\begin{tad}\label{tad:countable_cryptographic_scheme} A cryptographic scheme $(M,C,F)$ is called a {\em countable infinite cryptographic scheme}, if the sets $M$, $C$ and $F$ are
	countable infinite. If the set of ciphertexts $C'$ of a cryptographic scheme $(M',C',F')$ is finite then	$(M',C',F')$ is called a {\em finite cryptographic scheme}.
\end{tad}

The proof can be found in \cite{Cap21}.	

\section{Properties of Cryptographic Schemes}
\begin{dfn}\label{dfn:a-priori_and_a-posteriori_probabilities} If $(M,C,F)$ is a cryptographic 	scheme, then the probability of the occurrence of a message $m\in M$ is denoted with $\mu  (m)=P(m)$ and is called {\em a-priori probability} of the message $m$. Similarly the probability of the occurrence of the message $m$ under the condition that $m$ is mapped to $c$ by any $f\in F$ is denoted with $\nu (m,c)=P_{c=f(m)}(m)$ and is
	called {\em a-posteriori probability} of the message $m$ for a given ciphertext $c$.
\end{dfn}	
	
\begin{rem} In the case that a cryptographic system $(M,C,F)$ is finite, relative frequencies can be used to calculate the a-priori and a-posteriori probabilities. Then $\mu (m)=\frac{H(m)}{|M|}$ and $\nu (m,c)=\frac{\big|\{f\in F:c=f(m)\}\big|}{\sum_{m\in M}{\big|{f\in F:c=f(m)}\big|}}$. Here $H(m)$ means the frequency of the occurrence of the message $m$.
\end{rem}

\begin{dfn}\label{dfn:perfect_cryptographic_scheme}Let $(M,C,F)$ be a cryptographic scheme for
	which every $c\in C$ is a possible ciphertext and let $\mu$ and $\nu$ be its a-priori and a-posteriori probabilities.	Then $(M,C,F,\mu ,\nu )$ is called {\em perfect according to Shannon~\cite{Sha49}} as long as $\mu (m)=\nu (m,c)$ for any $m\in M$ and for any $c\in C$. 
\end{dfn}

\begin{rem} If $(M,C,F)$ is finite and if $\mu (m)=\mu (m_0)$ for any $m\in M$ then $(M,C,F,\mu ,\nu )$ is perfect as long as $\frac{1}{|M|}=\frac{\big|\{f\in F:c=f(m)\}\big|}{\sum_{m\in M}{\big|\{f\in F:c=f(m)\}\big|}}$ for any $m\in M$ and for any $c\in C$.
\end{rem}
	
\begin{dfn}\label{dfn:complete_cryptographic_scheme} Let $(M,C,F)$ be a cryptographic
	scheme with $M=\Z_2^r$ and $C=\Z_2^s$. Then the cryptographic transformation $f\in F$ is called {\em complete according to Kam and Davida~\cite{KD79}}, if there is at least one message $m_0= m_{0_1},m_{0_2},...,m_{0_r}\in M$ for every pair of indices $i\le r$ and $j\le
	s$, where a change in the $i$-th bit of $m_0\in M$ results in a change of the $j$-th bit of $c_0=c_{0_1},c_{0_2},...,c_{0_s}=f(m_0)\in C$. A cryptographic scheme $(M,C,F)$ that consists exclusively of	complete cryptographic transformations is called a {\em complete cryptographic scheme}.
\end{dfn}	
		
\section{Brief Introduction of the Geometry of the Laguerre Plane}
\begin{dfn}\label{dfn:laguerre_plane}
	The triple $(\boldsymbol{P},\boldsymbol{G},\boldsymbol{X})$ of {\em points} $\boldsymbol{P}$, {\em generators} $\boldsymbol{G}$ and {\em cycles} $\boldsymbol{X})$ is called {\em Laguerre plane}, if it satisfies following properties:
\begin{itemize}
	\item[{\em (L1)}] $\forall {\text {\em a}}\in \boldsymbol{P} {\exists}_1 [ \,{\text {\em a}}] \,\in \boldsymbol{G}:{\text {\em a}}\in [ \,{\text {\em a}}] \,$. $[ \,{\text {\em a}}] \,$ is the generator that passes through the point ${\text {\em a}}$.
	\item[{\em (L2)}] $\forall {\text {\em a}},{\text {\em b}},{\text {\em c}}\in \boldsymbol{P}$, with $[ \,{\text {\em a}}] \,\neq[ \,{\text {\em b}}] \,\neq[ \,{\text {\em c}}] \,\neq[ \,{\text {\em a}}] \,\exists_1{\text {\em C}}\in \boldsymbol{X}:{\text {\em a}},{\text {\em b}},{\text {\em c}}\in {\text {\em C}}$.
	\item[{\em (L3)}] $\forall {\text {\em G}}\in \boldsymbol{G},\forall {\text {\em C}}\in \boldsymbol{X}:|{\text {\em G}}\cap {\text {\em C}}|=1$.
    \item[{\em (L4)}] Touch axiom: $\forall {\text {\em C}}\in \boldsymbol{X}, {\text {\em a}}\in {\text {\em C}}$ and ${\text {\em b}}\in \boldsymbol{P}\setminus \big \{{\text {\em C}}\cup [{\text a}]\big \} {\exists}_1{\text {\em D}}\in \boldsymbol{X}: {\text {\em a}},{\text {\em b}}\in {\text {\em D}}$ and ${\text {\em C}}\cap {\text {\em D}}=\{{\text {\em a}}\}$.
    \item[{\em (L5)}] $\exists {\text {\em C}}\in  \boldsymbol{X}:|{\text {\em C}}|\geq 3$, there is $|\boldsymbol{P}|> 3$ and $\forall {\text {\em G}} \in \boldsymbol{G}:|{\text {\em G}}|\geq 2$.
\end{itemize}
\end{dfn}
	
	The Laguerre geometry can be defined as the points $\boldsymbol {P}$ on the surface of a circular cylinder in the Euclidian space. The cycles $\boldsymbol {X}$ of the Laguerre geometry are then the cone cuts of the circular cylinder with all planes that are not parallel to the axis of the cylinder. The tangents of the circular cylinder with planes touch the circular cylinder in lines, which are the generators of the Laguerre geometry. 
	
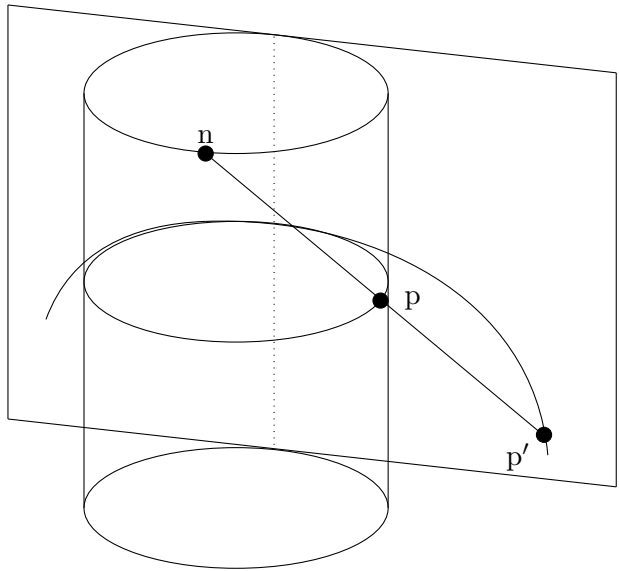
\begin{figure}
		\begin{center}
			\begin{tikzpicture}
			\draw (4.5,8) ellipse(2cm and 0.8cm);
			\draw (4.5,5.5) ellipse(2cm and 0.8cm);
			\draw (4.5,2.5) ellipse(2cm and 0.8cm);
			\draw (2.5,8) --(2.5,2.5);
			\draw (6.5,8) --(6.5,2.5);
			\draw (1.5,9.17) --(9.5,8.27);
			\draw (1.5,3.68) --(9.5,2.78);
			\draw (1.5,9.17) --(1.5,3.68);
			\draw (9.5,8.27) --(9.5,2.78);
			
	\draw (4.1,7.2) node [above] {${\text n}$}--(8.5,3.5) node [below left] {${\text p'}$};
	\draw [fill] (4.1,7.2) circle [radius=0.1]; 
	\draw [fill] (8.55,3.47) circle [radius=0.1]; 
	\draw [fill] (6.4,5.25) circle [radius=0.1] node [right] {${\hspace{5pt} \text p}$}; 
	\draw (2,5) to [out=70,in=182] (4.1,6.3);
		\draw (4.1,6.3) to [out=2,in=95] (8.6,3.2);
	\draw [dotted] (5,8.75) --(5,3.3);
			\end{tikzpicture}	
				\caption{Stereographic projection of the Laguerre geometry.}\label{fig:laguerre_plane}
		\end{center}
	\end{figure}

	The Laguerre geometry of the circular cylinder can be projected into the Euclidian plane by the use of a stereographic projection. The center of the projection is chosen as a point of the circular cylinder, the projection plane shall be in parallel to the axis of the cylinder and shall not contain the projection center. If the coordinate system of the Euclidian plane is chosen properly, the cycles of the Laguerre geometry that pass the center of the projection are mapped to the lines in the Euclidian plane that are not parallel to the $y$-axis. All other cycles of the cylinder are transferred to parabolas in the Euclidian plane with axes in parallel to the $y$-axis. The points of the Laguerre geometry that lie on the line ${\text L}_0$ passing the center of the projection are mapped to distant points. The set of distant points form the distant generator. Since ${\text L}_0$ is a line of the Euclidian space, the distant generator consists of as many points as a line of the Euclidian space. Figure~\ref{fig:laguerre_plane} illustrates the stereographic projection of the Laguerre geometry.
	
In order to find an analytical representation of the Laguerre plane, we want to introduce the {\em dual numbers}:

\begin{equation*}
	\mathbf{D}:=\{a+b\epsilon:a,b\in \mathbf{F}\}, \epsilon^2=0 \\
\end{equation*}
	
$\mathbf{D}$ is a local ring with $\mathbf{F}$ as a sub-field and $\mathbf{F}\epsilon$ as a maximum ideal. In $\mathbf{D}$ exists an involutorial automorphism $\overline{.\vphantom{G}}:\mathbf{D}\to \mathbf{D}$ that has the elements of $\mathbf{F}$ as fixpoints. We obtain the Laguerre plane by closing the dual numbers with a the distant generator. We will use the notation  $\overline{\mathbf{F}}:=\mathbf{F}\cup \{\infty\}$ for the closed field $\mathbf{F}$ and $\overline{\mathbf{D}}:=\mathbf{D}\cup \{\infty+\mathbf{F}\epsilon\}$ for the closure of the dual numbers.

\section{Describing the Laguerre Plane Using Equations}

With the right choice of the coordinate system, the cycles of the Laguerre plane are represented as parables and lines in $\overline{\mathbf{D}}$. A cycle ${\text C}\in \boldsymbol{X}$ is determined by the parameters $a,b,c \in \mathbf{F}$ with $(a,b)\neq (0,0)$ and is represented by following set:

\begin{equation*}
\big \{(x,y)\in \mathbf{F}^2:y=ax^2+bx+c \big \}\cup\{\infty-a\epsilon\}
\end{equation*}

\section{Describing the Laguerre Plane by the Use of Double Ratios}

A cycle in the Laguerre plane that includes the points ${\text a},{\text b},{\text c} \in \mathbf{D}$ with $[{\text a}] \,\neq[{\text b}] \,\neq[{\text c}] \,\neq[ {\text a}] \,$ can be represented with the following set:

\begin{equation*}
\Big \{{\text z}\in \mathbf{D}: [{\text z}] \, \neq [{\text a}] \,\text{and}\dr({\text a},{\text b},{\text c},{\text z}):=\nicefrac{\frac{{\text a}-{\text c}}{{\text a}-{\text z}}}{\frac{{\text b}-{\text c}}{{\text b}-{\text z}}}\in \mathbf{F}\Big \}\cup \{{\text a}\}\cup \Big \{\infty+\frac{{\text a}-{\text c}}{{\text b}-{\text c}}\epsilon\Big \}
\end{equation*}

\section{Describing the Laguerre Plane Using Fractional Linear Functions}
For the representation of the Laguerre plane with the use of fractional linear functions we will use a representation based on the closed dual numbers $\overline{\mathbf{D}}$ with coordinates of the field $\mathbf{F}$. Then the cycles in the Laguerre plane are determined as the pictures of a reference cycle, e.g. the set $\{z\in \overline{\mathbf{F}}\}$ regarding following mappings $\gamma$.

\begin{multline*}
\gamma:
z\mapsto \frac{{\text a}z+{\text b}}{{\text c}z+{\text d}},z\in \mathbf{F}, {\text a},{\text b},{\text c},{\text d} \in \mathbf{D},{\text a}{\text d}-{\text b}{\text c}\in \mathbf{D}\setminus \mathbf{F}\epsilon \\
\gamma(\infty)=\frac{{\text a}}{{\text c}}, \gamma (-\frac{{\text d}}{{\text c}})=\infty, \,\mbox {if} \, \frac{{\text d}}{{\text c}}\in \mathbf{F} \\
\end{multline*}

\section{Combinatorial Aspects of the Laguerre Plane}
\begin{figure}
	\begin{center}
		\begin{tikzpicture}
		\draw (0,1) --(6,1);
		\draw (1,0) node [below] {${\text A}$} --(1,5);
		\draw (3,0) node [below] {${\text B}$} --(3,5);
		\draw (2,1) node [below] {${\text C}$};
		\draw (0,4) to [out=275,in=210 ] (6,1.5); 
		\draw [fill] (1,2.05) circle [radius=0.1] node [below left] {${\text a}$}; 
		\draw [fill] (3,1.1) circle [radius=0.1] node [above right] {${\text b}$}; 
		
		\draw [fill] (4,1) circle [radius=0.1] node [below] {${\text c}$}; 
		\draw (6,1.5) node [right] {${\text D}_{\text a}$};

		\end{tikzpicture}	
		\caption{Bijection between two generators.}\label{fig:bijection_between_generators}
	\end{center}
\end{figure}
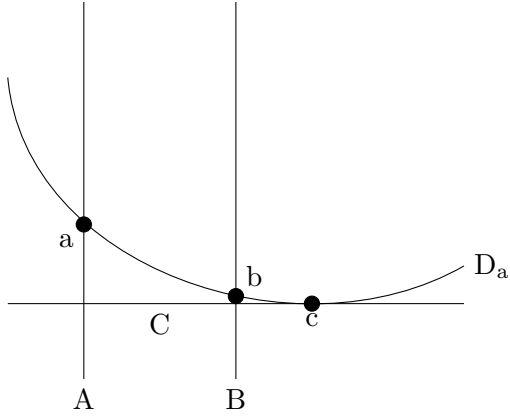

In the following we will focus on a Laguerre plane $(\boldsymbol{P},\boldsymbol{G},\boldsymbol{X})$ based on a finite field $\mathbf{F}$ with $|\mathbf{F}|=:q$. Let be ${\text C}\in \boldsymbol{X}$ with $|{\text C}|\geq 3$ (see (L5)), ${\text A}, {\text B}\in \boldsymbol{G}$ with ${\text A}\neq {\text B}$ and ${\text c}\in {\text C}\setminus( \,{\text A}\cup {\text B}) \,$. We assign every point ${\text a}\in {\text A}$ a cycle ${\text D}_{\text a}\in \boldsymbol{X}$ with ${\text a}\in {\text D}_{\text a}$ in the following way: If ${\text a}\in {\text C}$, then let ${\text D}_{\text a}={\text C}$. If ${\text a}\notin {\text C}$, then ${\text D}_{\text a}$ shall be uniquely defined by ${\text D}_{\text a} \cap {\text C}={\text c}$ (acc. to (L4)). Due to properties (L3) and (L4) the mapping 

\begin{equation*}
\gamma:\begin{cases}
	{\text A}\in {\text B}\\
	{\text a}\mapsto {\text D}_{\text a} \cap{\text B}
\end{cases}
\end{equation*} 

is bijective (see figure~\ref{fig:bijection_between_generators}).

Since the stereographic projection maps a generator of the Laguerre geometry $(\boldsymbol{P},\boldsymbol{G},\boldsymbol{X})$ to a line of the affine plane over the field $\mathbf{F}$, there are $q$ points on every generator. Furthermore there is $|q|=|{\text A}|=|{\text B}|$ for every ${\text A},{\text B}\in \boldsymbol{G}$. With (L3) follows $|\boldsymbol{P}|=|{\text A}|\cdot |{\text C}|$ and hence $|{\text C}|=|{\text D}|$ for every ${\text C},{\text D}\in \boldsymbol{X}$.

\begin{figure}
	\begin{center}
		\begin{tikzpicture}
		\draw (0,3) --(6,3) node [below] {${\text C}$};
		\draw (2,0) node [below] {${\text A}$} --(2,5);
		\draw (4.8,0) node [below] {${\text B}$} --(4.8,5);
		\draw (0.5,1) to [out=85,in=180 ] (6,4.5) node [right] {${\text E}$}; 
		\draw (1.5,0) to [out=88,in=137 ] (6,4) node [right] {${\text F}_{\text c}$};

		\draw [fill] (4.8,4.43) circle [radius=0.1] node [above left] {${\text b}$}; 
		\draw [fill] (2,1.8) circle [radius=0.1] node [right] {${\text c}$};

		\end{tikzpicture}	
		\caption{Bijection between two generators.}\label{fig:generator_to_cycle}
	\end{center}
\end{figure}
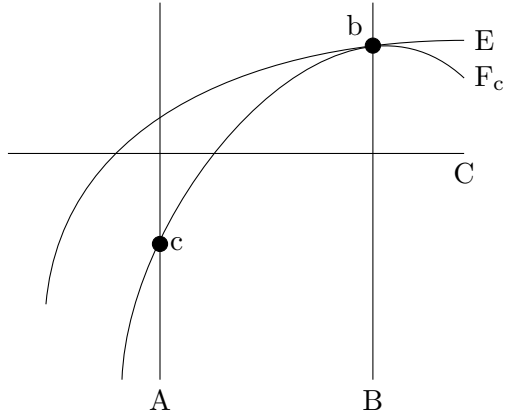

Let ${\text A},{\text B},{\text C}$ be as defined above, and let ${\text b}$ be a point with ${\text b}\in {\text B}\setminus{\text C}$ and ${\text E}\in \boldsymbol{X}$ be a cycle with ${\text b}\in {\text E}$. We assign every point ${\text c}\in {\text C}\setminus{\text B}$ the cycle ${\text F}_{\text c}\in \boldsymbol{X}$ with ${\text c}\in {\text F}_{\text c}$ and ${\text F}_{\text c}\cap{\text E}=\{{\text b}\}$. Basically this means that ${\text F}_{\text c}$ touches ${\text E}$ in ${\text b}$. If ${\text c}\in {\text E}$, then let ${\text F}_{\text c}:={\text C}$. Figure~\ref{fig:generator_to_cycle} illustrates this construction.

According to (L4) the cycle ${\text F}_{\text c}$ is uniquely determined by the cycle ${\text C}$ and the points ${\text b}$ and ${\text c}$. Because of (L3) the cycle ${\text F}_{\text c}$ has a common point with the generator ${\text A}$. Furthermore there is ${\text F}_{\text c}\cap{\text A}\neq{\text F}_{\text d}\cap{\text A}$ for any ${\text c}\neq{\text d}$. This provides a bijection from the set of points of ${\text C}\setminus \{{\text C}\cap {\text B}\}$ to ${\text A}$. Hence $|{\text C}|=|{\text A}|+1=q+1,\forall {\text C}\in \boldsymbol{X}$. Together with the results shown above we get $\boldsymbol{P}=q(q+1)$. Since a cycle is defined by exactly three points from three different generators (L2), we get together with (L1) as result for the number of cycles $|\boldsymbol{X}|=q^3$.

\section{Encryption in the Laguerre Plane}
In~\cite{Cap} an encryption scheme based on the M\"obius plane has been introduced. It is quite logical to transfer these results to the Laguerre plane. Again, we will use the incidence properties, in this case in the Laguerre plane, to create a mapping between three points representing message, cipher text and a key. To ensure that such an incidence exists, we choose messages, cipher texts and keys from different generators. Hence these points can never be parallel. 

We want to use a Laguerre plane over a finite field ${\mathbf F}$ with $|{\mathbf F}|\geq 3$ and hence $|\boldsymbol{G}|\geq 4$. So we select four generators ${\text G}_1, {\text G}_2, {\text G}_3, {\text G}_4$ of the Laguerre plane and assign them with the different sets of a cihper system. In this context we would like to use the notation $({\text a},{\text b},{\text c})^\circ$ for the cycle through the three points ${\text a},{\text b},{\text c}$ of the Laguerre plane which are pairwise not parallel. Also, as a small modification to definition~\ref{dfn:cryptographic_transformation} we don't expect the transformations $F$ of the cryptographic scheme to be reversible, but introduce a set of decryption transformations $G$ instead.

\begin{dfn}\label{dfn:laguerre_cipher}
	A cipher system $(M,C,F,G)$ is defined in the Laguerre plane $(\boldsymbol{P},\boldsymbol{G},\boldsymbol{X})$ as follows:

\begin{itemize}
\item[{\em Messages:}] $M:={\text {\em G}}_1$

\item[{\em Cipher texts:}] $C:={\text {\em G}}_2$

\item[{\em Keys:}] $K:={\text {\em G}}_3\times {\text {\em G}}_4$

\item[{\em Encryption functions:}]
$F:\begin{cases}
	K\times M\to C \\
	({\text {\em k}},{\text {\em l}},{\text {\em m}})\mapsto {\text {\em c}}:=({\text {\em k}},{\text {\em l}},{\text {\em c }})^\circ\cap {\text {\em G}}_2\\ 
\end{cases}$ 

\item[{\em Decryption functions:}]
$G:\begin{cases}
K\times C\to M \\
({\text {\em k}},{\text {\em l}},{\text {\em c}})\mapsto {\text {\em m}}:=({\text {\em k}},{\text {\em l}},{\text {\em c}})^\circ\cap {\text {\em G}}_1\\ 
\end{cases}$ 
\end{itemize}

The cipher system $(M,C,F,G)$ is called {\em Laguerre cipher}.
 
\end{dfn}

Due to property (L2) of the Laguerre plane there is exactly one cycle ${\text C}\in \boldsymbol{X}$ for any points ${\text m}\in M$ and ${\text k},{\text l}\in K$. Because of (L3) we know that $|{\text C}\cap C|=1$. Hence the cryptographic transformation $F$ is unique. Since the decryption function is defined analogue to the encryption function, it is also unique. 

\section{Cryptoanalysis of the Laguerre Cipher}

Let $(\boldsymbol{P},\boldsymbol{G},\boldsymbol{X})$ be a Laguerre plane over a finite field $\mathbf{F}$. To examine the property of perfectness we want to assume that all messages are equally distributed. The probability measures $\mu$ and $\nu$ shall be defined as described in definition~\ref{dfn:perfect_cryptographic_scheme}. The probability of the occurrence of a message ${\text m}\in M$ is $\mu({\text m})=\nicefrac{1}{|\mathbf{F}|}$, since there are $|\mathbf{F}|$ points on a generator. 

The probability $\nu({\text m},{\text c})$ that a message ${\text m}$ belongs to a certain ciphertext ${\text c}$ is determined by the number of elements in the set of cycles ${\text K}':=\{{\text C}\in \boldsymbol{X}:{\text m},{\text c}\in {\text C}\}$ that incide with the points ${\text m},{\text c}$ divided by the number of elements in the set of cycles ${\text K}'':=\{{\text C}\in \boldsymbol{X}:{\text m}\in {\text C}\}$ that incide with ${\text m}$. Since a generator ${\text G}'$ with ${\text m},{\text c}\notin {\text G}'$ has a unique common point with every cycle from ${\text K}'$ and since ${\text G}'$ has $|\mathbf{F}|$ points, we get $|{\text K}'|=|\mathbf{F}|$. The corresponding considerations about the number of cycles that pass ${\text m}$ lead to $|{\text K}''|=|\mathbf{F}|^2$. This leads to a probability for the occurrence of a message of $\mu({\text m})=\nicefrac{1}{|\mathbf{F}|}$ equaling the probability $\nu({\text m},{\text c})=\nicefrac{|\mathbf{F}|}{|\mathbf{F}|^2}$ that the message ${\text m}$ belongs to a certain ciphertext ${\text c}$.  Hence the Laguerre cipher fulfills the requirement of perfectness according to Shannon.

To examine the property of completeness according to Kam and Davida we define a Laguerre cipher as introduced in definition~\ref{dfn:laguerre_cipher} in a Laguerre plane  $(\boldsymbol{P},\boldsymbol{G},\boldsymbol{X})$ over a finite field $\mathbf{F}:=\Z_2^n$. The points in $\boldsymbol{P}$ have following coordinates:

\begin{equation*}
\begin{split}
{\text m}&=m_1+m_2\epsilon\\
{\text c}&=c_1+c_2\epsilon\\
{\text k}&=k_1+k_2\epsilon\\
{\text l}&=l_1+l_2\epsilon\\
\end{split}
\end{equation*}

We choose the generators of $(\boldsymbol{P},\boldsymbol{G},\boldsymbol{X})$ such that ${\text m}_1+\mathbf{F}\epsilon={\text E}_1$ and ${\text c}_1+\mathbf{F}\epsilon={\text E}_2$. Let also ${\text E}_1,{\text E}_2$ be in the finite. Otherwise the Laguerre plane can be transformed accordingly. Since the Laguerre cipher introduced in definition~\ref{dfn:laguerre_cipher} uses only points from ${\text E}_1$ as messages and only points from ${\text E}_2$ as ciphertexts, the cryptographic function $f:({\text m},{\text k},{\text l})\mapsto {\text c}$ effectively maps the 2nd coordinate $m_2$ of ${\text m}$ to the 2nd coordinate $c_2$ of ${\text c}$. We can describe these coordinages as binary vectors of length $n$. 

To show the property of completeness we select two indices $i,j\in {1,2,...,n}$ and define $e_i$ and $e_j$ as the respective unit vectors in $\Z_2^n$. For two points ${\text k}$ and ${\text l}$ on the generators ${\text E}_3$ and ${\text E}_4$ we have to find a ${\text m}$ and ${\text c}$ that hold following conditions:

\begin{equation}\label{cycle_m_c}
({\text k},{\text l},{\text m})^\circ \cap {\text E}_2=\{{\text c}\}
\end{equation}

\begin{equation}\label{cycle_mi_cj}
\begin{split}
({\text k},{\text l},{\text m}')^\circ \cap {\text E}_2&=\{{\text c}'\} \\{\text m}'&=m_1+(m_2+e_i)\epsilon \\ {\text c}'&=c_1+(c_2+e_j)\epsilon
\end{split}
\end{equation}

When we use double ratios to describe cycles in the Laguerre plane, the equations (\ref{cycle_m_c}) and (\ref{cycle_mi_cj}) can be presented as follows:

\begin{equation*}
\begin{split}
\dr({\text k},{\text l},{\text m},{\text c})&\in \mathbf{F}\\
\dr({\text k},{\text l},{\text m'},{\text c'})&\in \mathbf{F}
\end{split}
\end{equation*}

or

\begin{equation*}
\begin{split}
\frac{{\text k}-{\text m}}{{\text k}-{\text c}}&=c\cdot \frac{{\text l}-{\text m}}{{\text l}-{\text c}},c\in \mathbf{F}\\
\frac{{\text k}-{\text m}'}{{\text k}-{\text c}'}&=d\cdot \frac{{\text l}-{\text m}'}{{\text l}-{\text c}'},d\in \mathbf{F}
\end{split}
\end{equation*}

This leads to the following conditions for $m_2$, $c_2$, $c$ and $d$:

\begin{equation*}
\begin{aligned}
   (k_1-m_1)(l_1-c_1)&=c\cdot(k_1-c_1)(l_1-m_1) \\
    (k_1-m_1)(l_2-c_2)+(k_2-m_2)(l_1-c_1)&=c\cdot \big((k_1-c_1)(l_2-m_2)+(k_2-c_2)(l_1-m_1)\big) \\
    (k_1-m_1)(l_1-c_1)&=d\cdot(k_1-c_1)(l_1-m_1) \\
    (k_1-m_1)(l_2-c_2-e_j)&+(k_2-m_2-e_i)(l_1-c_1)= \\
= k\cdot \big((k_1-c_1)(l_2-m_2-e_i)&+(k_2-c_2-e_j)(l_1-m_1-e_i)\big)    
\end{aligned}
\end{equation*}

Values can be chosen for  ${\text m}_2$, ${\text c}_2$,  $c$ and $d$ to solve this systen of linear equations.

\section{Brief Introduction of the Geometry of the Minkowski Plane}\label{minkowski plane}

Another circle geometry is given with the Minkowski plane. Examples are hyperboloids in the three-dimensional projective space $\boldsymbol{P}$ over a commutative field $\mathbf{F}$. A hyperboloid can be represented as the set of points ${\text p}\in \boldsymbol{P}$, with ${\text p}:=\mathbf{F}^*(x_1,x_2,x_3,x_4)^\top$ and $(x_1,x_2,x_3,x_4)\in \big (\mathbf{F}^4 \big )^*$ that satisfy the quadratic equation $x_1x_4-x_2x_3=0$. In comparison to the circular cylinder of the Laguerre plane the hyperboloid of the Minkowski plane has two distinguished classes of straight lines $\boldsymbol{G}_1$ and $\boldsymbol{G}_2$ as generators. Two generators that are different from each other don't intersect, when they belong to the same class of generators, and they share exactly one common point, when they belong to different classes. The generator ${\text G}_i\in \boldsymbol{G}_i,i=1,2$ that contains the point ${\text p}$ is denoted with $[{\text p}]_i$. Two points ${\text p}_1$ and ${\text p}_2$ are called {\em connectable}, when $\left[{\text p_1}\right]_i\neq \left[{\text p_2}\right]_i$ for $i=1,2$, otherwise they are called {\em parallel}. The set of cycles of the Minkowski plane is determined by the figures that result from intersections with the planes of the three-dimensional projective space $\boldsymbol{P}$ that are no tangential planes of the hyperboloid. The following definition summarizes the properties of the Minkowski plane~\cite{KK88}.

\begin{dfn}\label{dfn:minkowski_plane}
	The sets of points $\boldsymbol{H}$, generators $\boldsymbol{G}_1\cup \boldsymbol{G}_2$ and cycles $\boldsymbol{X}$  constitute the {\em Minkowski  plane} $(\boldsymbol{H},\boldsymbol{G}_1\cup \boldsymbol{G}_2,\boldsymbol{X})$, if following properties are met:
	\begin{itemize}
		\item[{\em (N1)}] $\forall {\text {\em p}}\in \boldsymbol{H} {\exists}_2 {\text {\em G}}_i:{\text {\em G}}_i\in \boldsymbol{G}_i,[{\text {\em p}}]_i={\text G}_i, i \in \{1,2\}$
		\item[{\em (N2)}] $\forall {\text {\em G}}_1\in \boldsymbol{G}_1,\forall {\text {\em G}}_2\in \boldsymbol{G}_2 \exists_1 {\text {\em p}}\in \boldsymbol{H}:p\in {\text {\em G}}_i,i=1,2$
		\item[{\em (N3)}] $\forall {\text {\em G}}\in \boldsymbol{G}_1\cup \boldsymbol{G}_2:|G|\geq 2$
		\item[{\em (N4)}] $\forall {\text {\em G}}\in \boldsymbol{G}_1\cup \boldsymbol{G}_2,\forall {\text {\em C}}\in \boldsymbol{X}:|{\text {\em G}}\cap {\text {\em C}}|=1$
		\item[{\em (N5)}]For three pairwise connectable points ${\text {\em p}}_1,{\text {\em p}}_2,{\text {\em p}}_3 \in \boldsymbol{H}$ exists exactly one ${\text {\em C}}\in \boldsymbol{X}$ with ${\text {\em p}}_1,{\text {\em p}}_2,{\text {\em p}}_3 \in {\text {\em C}}$.
		\item[{\em (R)}] {\em Rectangle axiom:} For ${\text {\em A}},{\text {\em B}},{\text {\em C}}\in \boldsymbol{X}$ the set $\Big \{\big [[{\text {\em a}}]_1\cap {\text {\em B}}\big ]_2\cap \big [[{\text {\em a}}]_2\cap {\text {\em C}}\big ]_1:{\text {\em a}}\in {\text {\em A}}\Big \}$ is a cycle in $\boldsymbol{X}$.
		\item [{\em (T)}] {\em Touch axiom:} $\forall {\text {\em A}}\in \boldsymbol{X}, {\text {\em a}}\in {\text {\em A}}, {\text {\em b}}\in \boldsymbol{H}\setminus \big({\text {\em A}}\cup [a]_1\cup [a]_2\big) {\exists}_1{\text {\em B}}\in \boldsymbol{X}: {\text {\em b}}\in {\text {\em B}}, {\text {\em A}}\cap {\text {\em B}}=\{{\text {\em a}}\}$
		\item[{\em (S)}] {\em Symmetry axiom:} If ${\text {\em C}},{\text {\em D}}\in \boldsymbol{X}$ are two cycles and, if the point ${\text {\em p}}\in {\text {\em C}}\setminus {\text {\em D}}$ fulfills the condition $\big [[{\text {\em p}}]_1 \cap {\text {\em D}}\big ]_2 \cap \big [[{\text {\em p}}]_2 \cap {\text {\em D}}\big ]_1 \in {\text {\em C}}$ then $\big [[{\text {\em x}}]_1 \cap {\text {\em D}}\big ]_2 \cap \big [[{\text {\em x}}]_2 \cap {\text {\em D}}\big ]_1 \in {\text {\em C}}$ holds for any ${\text {\em x}}\in {\text {\em C}}$. 
	\end{itemize}
\end{dfn}

\begin{figure}
	\begin{center}
		\begin{tikzpicture}
		\draw (-2,8) --(7,0);
		\draw (7,0) --(15,3);
		\draw (-2,8) --(6,10);
		\draw [dashed](6,10) --(15,3);
		\draw (6,10) --(8,8.44);
		\draw (11.4,5.8)--(15,3);
		\draw (1,5.5) ellipse(1.5cm and 3cm);
		\draw[dashed] (10,7) ellipse(1.5cm and 3cm);
		\draw (1,8.5) to [out=-35,in=225] (10,10);
		\draw (2.35,4.19) to [out=45,in=177](11.4,5.8);
		\draw (2.5,5) to [out=45,in=177] (11.5,6.6);
		\draw (6,1.5) to [out=80,in=160] (12,3);
		\draw [dashed] (4.4,6) ellipse (0.7cm and 1.4cm);
		\draw [dashed](1,2.5) to [out=45,in=155] (10,4);
		\draw [dashed](-0.4,6.55) to [out=-10,in=210] (8.5,7.95);
		\draw (4.4,7.4) node [above] {${\text n}$}--(8.5,3.5) node [below left] {${\text p'}$};
		\draw [fill] (4.4,7.4) circle [radius=0.1]; 
		\draw [fill] (8.55,3.47) circle [radius=0.1]; 
		\draw [fill] (5.27,6.6) circle [radius=0.1] node [above right] {${\text p}$}; 
		\end{tikzpicture}	
		\caption{Stereographic projection of the Minkowski geometry.}\label{fig:minkowski_plane}
	\end{center}
\end{figure}
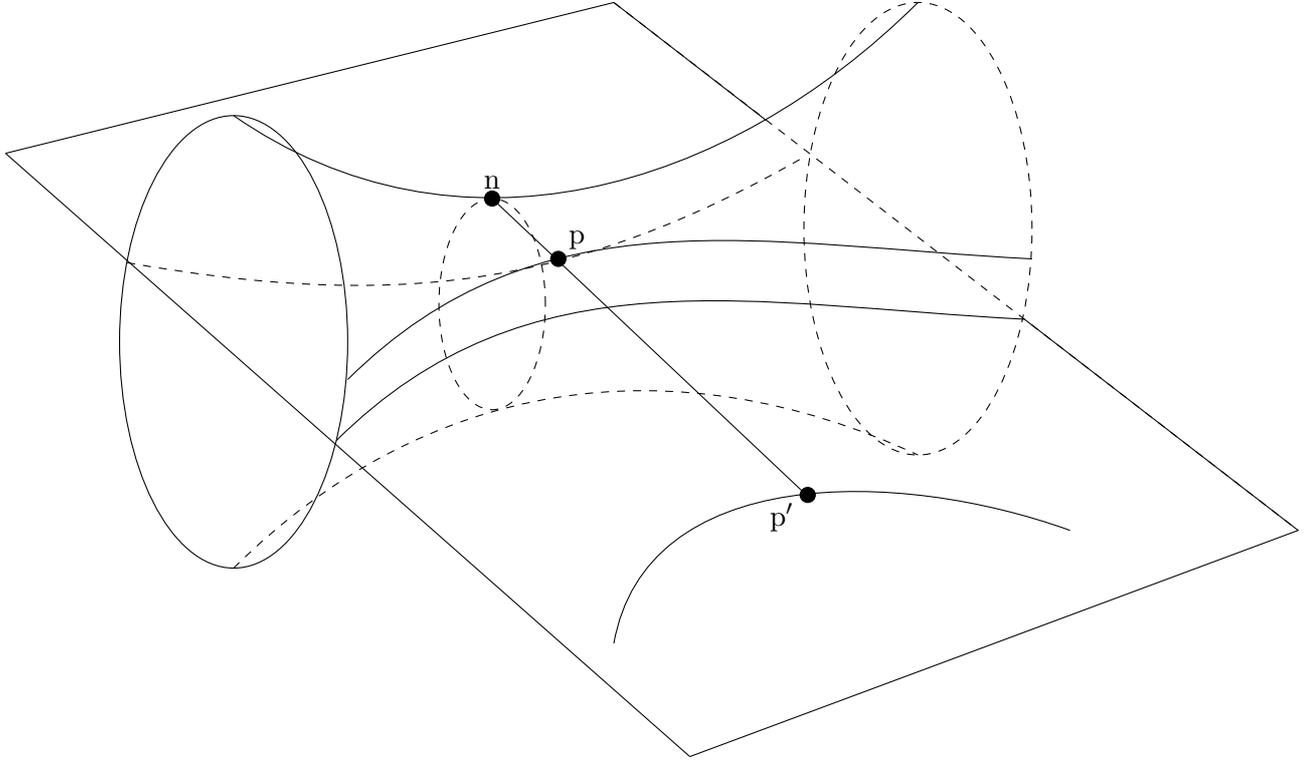

The hyperboloid $\boldsymbol{H}$ can be projected into the Euclidean plane $\R^2$ by means of a central projection. Therefore a point ${\text n}\in \boldsymbol{H}$ is chosen as the center of the projection and a plane ${\text P}$ of the projective space with ${\text n}\notin {\text P}$ as projection plane. If ${\text T}_{\text n}$ is the tangential plane in the point ${\text n}$, we consider $\overset{\circ} {\text E} :={\text E}\setminus {\text T}_{\text n}$. The generators of $\boldsymbol{H}$ are mapped to two sets of parallel lines in $\overset{\circ}{\text E}$. The cycles ${\text C}\in \boldsymbol{X}$ that do not pass the point ${\text n}$ are transformed to hyperbolas with the images of the generators in $\boldsymbol{H}$ as asymptotes. The cycles in $\boldsymbol{X}$ that pass the projection center ${\text n}$ are mapped to the lines of $\overset{\circ} {\text E}$ that are different from the images of the two sets of generators. Figure \ref{fig:minkowski_plane} visualizes this projection.

Let $\mathbf{F}$ be the field of the coordinates of the three-dimensional projective space $\boldsymbol{P}$. The points of this projective space can be written as 2x2-matrices ${\text M}_{22}$ over the field $\mathbf{F}$. The set of points can be written as $\boldsymbol{P}:=\{\mathbf{F}^*{\text x}:{\text x}\in {\text M}_{22}\}$. The subset $\boldsymbol{H}:=\{\mathbf{F}^*{\text x}:\determ {\text x}=0\}$ defines a hyperboloid in $\boldsymbol{P}$. 

For every point $\mathbf{F}^*{\text a}\in \boldsymbol{P}$ we define $(\mathbf{F}^*{\text a})^\perp:=\{\mathbf{F}^*{\text x}\in \boldsymbol{P}:\determ ({\text x}+{\text a})=\determ {\text x} + \determ {\text a}\}$ as the polar plane of the point $\mathbf{F}^*{\text a}$ regarding the quadric $\boldsymbol{H}$. The plane $(\mathbf{F}^*{\text a})^\perp$ is a tangential plane of $\boldsymbol{H}$ only, if $\determ {\text a}=0$. Since every plane of the projective space is a polar plane for some point in the projective space, the sets $(\mathbf{F}^*{\text a})^\perp \cap \boldsymbol{H}$ with $\determ a \neq 0$ determine all sections of $\boldsymbol{H}$ with planes that are not tangential.

Concluding, the Minkowski plane $(\boldsymbol{H},\boldsymbol{G}_1\cup \boldsymbol{G}_1,\boldsymbol{X})$ over the commutative field $\mathbf{F}$  can be represented with following coordinates:

\begin{equation*}
\begin{split}
	\boldsymbol{H} & = \big \{\mathbf{F}^*{\text x}:\determ {\text x}=0  \big \} \\
	\boldsymbol{G}_1 & = \left \{ \left \{ \mathbf{F}^* \left ( \begin{array}{cc}
	g_1x_1 & g_2x_1 \\
	g_1x_2 & g_2x_2
	\end{array} \right ):(x_1,x_2)\in {\mathbf{F}^2}^* \right \}:(g_1,g_2)\in {\mathbf{F}^2}^* \right \} \\
	\boldsymbol{G}_2 & = \left \{ \left \{ \mathbf{F}^* \left ( \begin{array}{cc}
x_1g_1 & x_2g_1 \\
x_1g_2 & x_2g_2
\end{array} \right ):(x_1,x_2)\in {\mathbf{F}^2}^* \right \}:(g_1,g_2)\in {\mathbf{F}^2}^* \right \} \\
	\boldsymbol{X} & = \big \{ (\mathbf{F}^*{\text a})^\perp \cap \boldsymbol{H}:{\text a} \in {\text M}_{22},\determ {\text a}\neq 0\big \}
\end{split}
\end{equation*}

\section{Describing the Minkowski Plane using Equations}
The set ${\text Q}$ defines a one-sheeted hyperboloid in the three-dimensional real projective space $\boldsymbol{P}$.

\begin{equation*}
{\text Q}:=\big \{(x,y,z,t)\in \boldsymbol{P}:x^2-y^2+z^2-t^2=0  \big \}
\end{equation*}

The stereographic projection $\sigma$ with projection center ${\text n}=\mathbf{F}^*(0,0,1,1)$ maps ${\text Q}$ to the plane ${\text P}:z=0$. The generators of the Minkowski plane are mapped to two crossing bundles of parallel lines. If ${\text T}_{\text n}$ is the tangent plane to the hyperboloid ${\text Q}$ through the point ${\text n}$, then ${\text T}_{\text n}$  and ${\text Q}$ intersect in two lines that meet in the point ${\text n}$. These two lines have no image in the plane ${\text P}$. By this reason the projection $\sigma$ is analytically continued by identifying the image of these two lines with the two distant generators of the Minkowski plane. 

The cycles of the Minkowski plane can be described in the plane ${\text P}$ by replacing the Euclidean metric that is defined by the quadratic form $x^2+y^2$ with the indefinite form $x^2-y^2$. If a suitable coordinate system is chosen for ${\text P}$, a cycle ${\text C}$ can be described in the finit as the following set of points:

\begin{equation} \label{minkowski_cycle}
{\text C}=\big \{(x,y)\in \R^2:(x-a)(y-b)=c,a,b,c\in \R   \big \}
\end{equation}

Here the cycles of the Minkowski plane are of the shape of hyperbolas with asymptotes which are parallel to the lines $y=c$ and $x=c$ with $c\in \R$. 

This model of the Minkowski plane can be generalized by using a field $\mathbf{F}$ instead of the real numbers $\R$. In this case the cycles of the Minkowski plane fulfill following condition:

\begin{equation} \label{minkowski_cycle}
{\text C}=\big \{(x,y)\in \mathbf{F}^2:(x-a)(y-b)=c,a,b,c\in \mathbf{F}   \big \}
\end{equation}

\section{Combinatorial Aspects of the Minkowski Plane}
We consider a Minkowski plane based on a finite field $\mathbf{F}$ with $|\mathbf{F}|=:q$ elements. From the analytical representation of the Minkowski plane introduced in section~\ref{minkowski plane} follows for the set of generators $\overline{\boldsymbol{G}_1}\cup \overline{\boldsymbol{G}_2}$ that $|\overline{\boldsymbol{G}_i}|=q+1$ for $i=1,2$. With (N2) follows for the number of points in the Minkowski plane $|\boldsymbol{H}|=(q+1)^2$. Furthermore we gain the number of cycles $|\boldsymbol{X}|=(q+1)q(q-1)$ from properties (N1), (N2), (N4) and (N5).

\section{Encryption based on the Minkowski Plane}
\begin{figure}
	\begin{center}
		\begin{tikzpicture}
		\draw (1,0) node [right] {$\left[{\text k}\right]_1$}--(1,5.5);
		\draw (0.5,5) --(6,5) node [right] {$\left[{\text k} \right]_2$};
		\draw (0,0.5) to [out=3,in=267] (5.5,6) node [right] {${\text C}$};
		\draw [fill] (1,5) circle [radius=0.1] node [above left] {${\text k}$}; 
		\draw [fill] (1,0.65) circle [radius=0.1] node [above left] {${\text m}$}; 
		\draw [fill] (5.35,5) circle [radius=0.1] node [above left] {${\text c}$}; 
		\end{tikzpicture}	
		\caption{Encryption in the Minkowski plane with parallel key.}\label{fig:encryption_with_parallel_key}
	\end{center}
\end{figure}
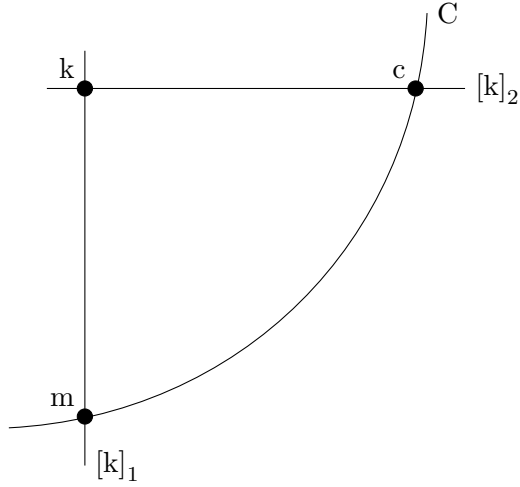

The encryption scheme introduced in~\cite{Cap21} for the M\"obius plane can be transferred to the Minkowski plane $(\boldsymbol{H},\boldsymbol{G}_1\cup \boldsymbol{G}_2,\boldsymbol{X})$. Messages and cipher texts can be identified with the set of triples of points in 
$\boldsymbol{H}$. Here we have to consider the case that the point ${\text m}$ representing the message and the point ${\text k}$ representing the key are parallel. Then these two points are not connectable. In this case the cipher text can be determined using following construction: Lets consider that $\left[{\text m} \right]_1=\left[{\text k}\right]_1$, then we define the associated ciphertext with $c:={\text M}\cap \left[{\text k}\right]_2$, where ${\text M}$ is determined by the three messages ${\text m},{\text n},{\text o}$. According to (N4) there is always such a point of intersection (see figure~\ref{fig:encryption_with_parallel_key}). The corresponding practice can be applied in case of $\left[{\text m} \right]_2=\left[{\text k}\right]_2$.

With these preparations we can define a ciphersystem in the Minkowski plane. Therefore a point $\infty$ shall be distinguished in $(\boldsymbol{H},\boldsymbol{G}_1\cup \boldsymbol{G}_2,\boldsymbol{X})$. Then we consider the derivation of the Minkowski plane in that point:

\begin{equation*}
\begin{split}
	\boldsymbol{H}^\infty&:=\boldsymbol{H} \setminus \big( \left[\infty \right]_1 \cup \left[\infty \right]_2 \big) \\
	\boldsymbol{X}^\infty&:=\Big\{{\text C}\setminus\{\infty\}:{\text C}\in \boldsymbol{X}(\infty), \boldsymbol{X}(\infty):=\big \{{\text C\in \boldsymbol{X}:\infty\in {\text C}}  \big \}\Big\} \\
	\boldsymbol{G}_1^\infty&:=\Big \{{\text G}\setminus \left[\infty \right]_2:{\text G}\in \boldsymbol{G}_1\setminus \big \{\left[\infty \right]_1\big \} \Big \} \\
	\boldsymbol{G}_2^\infty&:=\Big \{{\text G}\setminus \left[\infty \right]_1:{\text G}\in \boldsymbol{G}_2\setminus \big \{\left[\infty \right]_2\big \} \Big \} \\
\end{split}
\end{equation*}

Then $(\boldsymbol{H}^\infty,\boldsymbol{G}_1^\infty\cup \boldsymbol{G}_2^\infty\cup \boldsymbol{X}^\infty)$ is an affine plane. For two given points ${\text a},{\text b}\in \boldsymbol{H}^\infty$ with ${\text a}\neq {\text b}$ we denote $\overline{{\text a},{\text b}}$ as the straight line uniquely determined by these two points. Furthermore the sets of degenerated, ordinary and all hyperbolas in the affine plane shall be defined as follows:

\begin{equation*}
\begin{split}
	[{\text x}]&:=\left[{\text x}\right]_1\cup \left[{\text x}\right]_2 \forall {\text x}\in \boldsymbol{H}^\infty\\
	\big[\boldsymbol{H}^\infty\big]&:=\big \{[{\text x}]\setminus [\infty]:{\text x\in \boldsymbol{H}^\infty} \big \} \\
	\boldsymbol{X}_\infty&:=\big\{{\text X}\setminus[\infty]:{\text X}\in \boldsymbol{X}\setminus \boldsymbol{X}(\infty) \big\} \\
	\overline{\boldsymbol{X}}&:=\boldsymbol{X}_\infty\cup \big[\boldsymbol{H}^\infty\big] \\
\end{split}
\end{equation*}

With this preparation we can define a cipher system in the Minkowski plane. For this cipher we will use straight lines as keys, however the following definition can be generalized in a way that cycles of the Minkowski plane are used as keys~\cite{Cap}.

\begin{dfn}\label{dfn:minkowski_cipher}
Let $(\boldsymbol{H}^\infty,\boldsymbol{G}_1^\infty\cup \boldsymbol{G}_2^\infty\cup \boldsymbol{X}^\infty)$ be the derivation of the Minkowski plane $(\boldsymbol{H},\boldsymbol{G}_1\cup \boldsymbol{G}_2,\boldsymbol{X})$ in the point $\infty$ and let ${\text {\em m}}_1,{\text {\em m}}_2,{\text {\em m}}_3\in \boldsymbol{H}^\infty$ be three different points of the affine plane that cannot be connected with one straight line. Then we want to make following definitions:
\begin{itemize}
\item [{\em Messages:}] $\left(\begin{array}{c}\boldsymbol{H}^\infty \\ 3 \\ \end{array}\right)$ shall denote the set of triples of points in $\boldsymbol{H}^\infty$. Then the set of messages can be defined as follows:

  	$M:=\left\{({\text {\em m}}_1,{\text {\em m}}_2,{\text {\em m}}_3)\in \left(\begin{array}{c}\boldsymbol{H}^\infty \\ 3 \\ \end{array}\right): \begin{array}{l}
  	\left[{\text {\em m}}_1\right]\neq\left[{\text {\em m}}_2\right]\neq\left[{\text {\em m}}_3\right]\neq\left[{\text {\em m}}_1\right] \\ \nexists {\text {\em L}}\in \boldsymbol{X}^\infty:\{{\text {\em m}}_1,{\text {\em m}}_2,{\text {\em m}}_3\}\in {\text {\em L}} \\
  	\end{array}	\right\} $

\item [{\em Cipher texts:}] $C:=M$ 

\item [{\em Keys:}] $K:=\big \{({\text {\em k}}_1,{\text {\em k}}_2,{\text {\em k}}_3)\in \left(\boldsymbol{H}^\infty \right)^3 \big \}$ with following properties:

	{\renewcommand \labelitemi{}
	\begin{itemize}[label={}]
		\item  ${\text {\em k}}_i\notin({\text {\em m}}_1,{\text {\em m}}_2,{\text {\em m}}_3)^\circ$ and
		\item $({\text {\em m}}_1,{\text {\em m}}_2,{\text {\em m}}_3)^\circ\cap\overline{{\text {\em m}}_i,{\text {\em k}}_i}\cap\overline{{\text {\em m}}_j,{\text {\em k}}_j}=\varnothing$ for $i,j\in\{1,2,3\}$ and $i\neq j$ 
	\end{itemize}
}
\item [{\em Encryption Funtion:}]$F:K\times M\to C$
\item [{\em Encryption:}]
	${\text {\em c}}_i:=\begin{cases}
	\overline{{\text {\em m}}_i,{\text {\em k}}_i}\cap ({\text {\em m}}_1,{\text {\em m}}_2,{\text {\em m}}_3)^\circ \setminus \{{\text {\em m}}_i\},&{\text if} \big |\overline{{\text {\em m}}_i,{\text {\em k}}_i}\cap ({\text {\em m}}_1,{\text {\em m}}_2,{\text {\em m}}_3)^\circ \big |=2 \\ &{\text and} \left[{\text {\em m}}_i\right]\cap \left[{\text {\em k}}_i\right] \leq 2 \\
	{\text {\em m}}_i, &{\text if} \big |\overline{{\text {\em m}}_i,{\text {\em k}}_i}\cap ({\text {\em m}}_1,{\text {\em m}}_2,{\text {\em m}}_3)^\circ \big |=1 \\ &{\text and} \left[{\text {\em m}}_i\right]\cap \left[{\text {\em k}}_i\right] \leq 2 \\
	\left[{\text {\em k}}_i\right]_2\cap ({\text {\em m}}_1,{\text {\em m}}_2,{\text {\em m}}_3)^\circ, &{\text if} \left[{\text {\em m}}_i\right]_1=\left[{\text {\em k}}_i\right]_1\\ 
	\left[{\text {\em k}}_i\right]_1\cap ({\text {\em m}}_1,{\text {\em m}}_2,{\text {\em m}}_3)^\circ, &{\text if} \left[{\text {\em m}}_i\right]_2=\left[{\text {\em k}}_i\right]_2\\ 
	\end{cases}$ 
\item [{\em Decryption Funtion:}]$G:K\times C\to M$	
\item[{\em Decryption:}]
	${\text {\em m}}_i:=\begin{cases}
\overline{{\text {\em c}}_i,{\text {\em k}}_i}\cap ({\text {\em c}}_1,{\text {\em c}}_2,{\text {\em c}}_3)^\circ \setminus \{{\text {\em c}}_i\},&{\text if} \big |\overline{{\text {\em c}}_i,{\text {\em k}}_i}\cap ({\text {\em c}}_1,{\text {\em c}}_2,{\text {\em c}}_3)^\circ \big |=2 \\ &{\text and} \left[{\text {\em c}}_i\right]\cap \left[{\text {\em k}}_i\right] \leq 2 \\
{\text {\em c}}_i, &{\text if} \big |\overline{{\text {\em c}}_i,{\text {\em k}}_i}\cap ({\text {\em c}}_1,{\text {\em c}}_2,{\text {\em c}}_3)^\circ \big |=1 \\ &{\text and} \left[{\text {\em c}}_i\right]\cap \left[{\text {\em k}}_i\right] \leq 2 \\
\left[{\text {\em k}}_i\right]_2\cap ({\text {\em c}}_1,{\text {\em c}}_2,{\text {\em c}}_3)^\circ, &{\text if } \left[{\text {\em c}}_i\right]_1=\left[{\text {\em k}}_i\right]_1\\ 
\left[{\text {\em k}}_i\right]_1\cap ({\text {\em c}}_1,{\text {\em c}}_2,{\text {\em c}}_3)^\circ, &{\text if} \left[{\text {\em c}}_i\right]_2=\left[{\text {\em k}}_i\right]_2\\ 
\end{cases}$ 

\end{itemize}

	This cipher system is called {\em Minkowski cipher}.
	
\end{dfn}

\section{Cryptoanalysis of the Minkowski Cipher}

To analyze the property of perfectness according to definition~\ref{dfn:perfect_cryptographic_scheme} for the Minkowski cipher, we consider the Minkowski plane $(\boldsymbol{H},\boldsymbol{G}_1\cup \boldsymbol{G}_2,\boldsymbol{X})$ over a finite field $\mathbf{F}$ with $q:=|\mathbf{F}|$ elements. When we select a point $\infty\in \boldsymbol{X}$, we get an affine plane $(\boldsymbol{H}^\infty,\boldsymbol{G}_1^\infty\cup \boldsymbol{G}_2^\infty\cup\boldsymbol{X}^\infty)$ as the derivation of the Minkowski plane in the point $\infty$. Let $({\text m}_1,{\text m}_2,{\text m}_3)\in M$ be a message from the set of messages as defined in definition~\ref{dfn:minkowski_cipher}. Then there exists exactly one cycle $({\text m}_1,{\text m}_2,{\text m}_3)^\circ\in \boldsymbol{X}_\infty$. 

To show the perfectness of the Minkowski cipher, we look at the encryption of the message points ${\text m}_1,{\text m}_2,{\text m}_3$ that all lie on the same cycle ${\text C}$ of the Minkowski plane. Probability measures $\mu$ and $\nu$ as introduced in definition~\ref{dfn:a-priori_and_a-posteriori_probabilities}  shall be defined. We will further assume that all messages occur with the same probability. The number of possible keys that can be used to encrypt ${\text m}_i$ to ${\text c}_i$ is determined by the conditions for the set of keys $K$ as defined in definition~\ref{dfn:minkowski_cipher}. 

We look at the encryption of the message point ${\text m}_1$. The message $({\text m}_1,{\text m}_2,{\text m}_3)$ determines the hyperbola ${\text H}:=({\text m}_1,{\text m}_2,{\text m}_3)^\circ\in\boldsymbol{X}_\infty$. The message point ${\text m}_1$ can be one of the $q-1$ points of the hyperbola ${\text H}$. Hence the probability for the occurrence of ${\text m}_1$ is $\mu({\text m}_1)=\nicefrac{1}{(q-1)}$. 

The set $K_1$ of keys of the Minkowski cipher that can be used to encrypt ${\text m}_1$ is described by the set of all points in the affine plane that are not on the hyperbola ${\text H}$, i.e. $\boldsymbol{X}_\infty\setminus {\text H}$. Hence $|{\text K}_1|=q^2-(q-1)$. To find the number of keys that can encrypt the message ${\text m}_1$ to the cipher text ${\text c}_1$ we have to distinguish two cases.

\begin{itemize}
	\item [${\text m}_1={\text c}_1$] Every point of the tangent to the hyperbola ${\text H}$ through the point ${\text c}_1$ except the point ${\text c}_1$ itself is a possible key. Hence $\big |\{f\in F:({\text m}_1,{\text c}_1)\subset f\}\big |=q-1$.
	\item[{\em ${\text m}_1\neq{\text c}_1$}] All points of the line $\overline{\vphantom{l}{\text {\em m}}_1,{\text {\em c}}_1}$ except ${\text m}_1$ and ${\text c}_1$ and the points $\left[{\text m}_1\right]_1\cap \left[{\text c}_1\right]_2$ and $\left[{\text m}_1\right]_2\cap \left[{\text c}_1\right]_1$ are possible keys. Hence $\big |\{f\in F:({\text m}_1,{\text c}_1)\subset f\}\big |=q$. 
\end{itemize}

Altogether we found in the case ${\text m}_1={\text c}_1$ the a-posteriori probability $\nu({\text m}_1,{\text c}_1)=\frac{q-1}{q^2-q+1}$ and in the case ${\text m}_1\neq {\text c}_1$ the a-posteriori probability $\nu({\text m}_1,{\text c}_1)=\frac{q}{q^2-q+1}$. As the a-priori probability for the encryption of ${\text m}_1$ holds $\mu({\text m}_1)=\frac{1}{q-1}$ we can conclude that the encryption of the first message point ${\text m}_1$ is perfect in first approximation for large $q$. 

We continue with the analysis of the encryption of the message point ${\text m}_2$. Since the three message points ${\text m}_1,{\text m}_2,{\text m}_3$ determine a unique cycle in $\boldsymbol{X}_\infty$, we know that ${\text m}_2\neq {\text m}_1$. Hence we obtain an a-priori probability of $\mu({\text m}_2)=\nicefrac{1}{(q-2)}$. Due to the condition given for the selection of keys in definition~\ref*{dfn:minkowski_cipher}, the set of keys $K_2$ for the encryption of ${\text m}_2$ is reduced  about those keys that would map ${\text m}_2$ to ${\text m}_1$ or ${\text c}_1$. Again the two cases ${\text m}_1={\text c}_1$ and ${\text m}_1\neq{\text c}_1$ need to be distinguished. In case of ${\text m}_1={\text c}_1$ there is $K_2=K_1\setminus \left \{\overline{\vphantom{l}{\text {\em m}}_1,{\text {\em m}}_2}\right \}$ and hence $|K_2|=q^2-q+1-q=(q-1)^2$. In case of ${\text m}_1\neq {\text c}_1$ we gain $K_2=K_1\setminus \left \{\overline{\vphantom{l}{\text {\em m}}_1,{\text {\em m}}_2}, \overline{\vphantom{l}{\text {\em m}}_2,{\text {\em c}}_1} \right \}$ leading to $|K_2|=q^2-q+1-q-q=q^2-3q+1$. 

For the determination of the a-posteriori probabilities for the encryption of ${\text m}_2$ the cases ${\text m}_2={\text c}_2$ and ${\text m}_2\neq {\text c}_2$ have to be distinguished:  

\begin{itemize}
	\item [${\text m}_2={\text c}_2$:] Every point of the tangent to the hyperbola ${\text H}$ through the point ${\text c}_2$ is a possible key. Hence $\big | \big \{f\in F:({\text m}_2,{\text c}_2\subset f)  \big \}\big |=q-1$.
	\item [${\text m}_2\neq {\text c}_2$:] Any point of the line through ${\text m}_2$ and ${\text c}_2$ except the two points ${\text m}_2$ and ${\text c}_2$ themselves and the points $\left[{\text m}_2\right]_1\cap \left[{\text c}_2\right]_2$ and $\left[{\text m}_2\right]_2\cap \left[{\text c}_2\right]_1$ are possible keys. Hence $\big | \big \{f\in F:({\text m}_2,{\text c}_2\subset f)  \big \}\big |=q$.
\end{itemize}

The following table summarizes the resulting a-priori and a-posteriori probabilities for the encryption of ${\text m}_2$:

\begin{center}
	\begin{tabular}{|c|c|c|}
		\hline 
		Case & A-priori probability & A-posteriori probability \\
		\hline
		${\text m}_1= {\text c}_1$,${\text m}_2= {\text c}_2$ & $\mu({\text m}_2)=\frac{1}{q-2}$ & $\nu ({\text m}_2,{\text c}_2)=\frac{1}{q-1}$ \\
		\hline
		${\text m}_1= {\text c}_1$,${\text m}_2\neq {\text c}_2$ & $\mu({\text m}_2)=\frac{1}{q-2}$ & $\nu ({\text m}_2,{\text c}_2)=\frac{q}{(q-1)^2}$ \\
		\hline
		${\text m}_1\neq {\text c}_1$,${\text m}_2= {\text c}_2$ & $\mu({\text m}_2)=\frac{1}{q-2}$ & $\nu ({\text m}_2,{\text c}_2)=\frac{q-1}{q^2-3q+1}$ \\
		\hline
		${\text m}_1\neq {\text c}_1$,${\text m}_2\neq {\text c}_2$ & $\mu({\text m}_2)=\frac{1}{q-2}$ & $\nu ({\text m}_2,{\text c}_2)=\frac{q}{q^2-3q+1}$ \\
		\hline
	\end{tabular}
\end{center}

Hence the encryption of the message point ${\text m}_2$ is also approximately perfect for large $q$.

The a-priori probability for the encryption of ${\text m}_3$ is $\mu({\text m}_3)=\nicefrac{1}{(q-3)}$. The following table shows also the a-posteriori probabilities:

\begin{center}
	\begin{tabular}{|c|c|c|}
		\hline 
		Case & A-priori probability & A-posteriori probability \\
		\hline
		${\text m}_1= {\text c}_1$,${\text m}_2= {\text c}_2$,${\text m}_3= {\text c}_3$ & $\mu({\text m}_2)=\frac{1}{q-3}$ & $\nu ({\text m}_3,{\text c}_3)=\frac{q-1}{q^2-3q+1}$ \\
		\hline
		${\text m}_1= {\text c}_1$,${\text m}_2= {\text c}_2$,${\text m}_3\neq {\text c}_3$ & $\mu({\text m}_2)=\frac{1}{q-3}$ & $\nu ({\text m}_3,{\text c}_3)=\frac{q}{q^2-3q+1}$ \\
		\hline
		${\text m}_1= {\text c}_1$,${\text m}_2\neq {\text c}_2$,${\text m}_3= {\text c}_3$ & $\mu({\text m}_2)=\frac{1}{q-3}$ & $\nu ({\text m}_3,{\text c}_3)=\frac{q-1}{q^2-4q+1}$ \\
\hline
		${\text m}_1= {\text c}_1$,${\text m}_2\neq {\text c}_2$,${\text m}_3\neq {\text c}_3$ & $\mu({\text m}_2)=\frac{1}{q-3}$ & $\nu ({\text m}_3,{\text c}_3)=\frac{q}{q^2-4q+1}$ \\
\hline
		${\text m}_1\neq {\text c}_1$,${\text m}_2= {\text c}_2$,${\text m}_3= {\text c}_3$ & $\mu({\text m}_2)=\frac{1}{q-3}$ & $\nu ({\text m}_3,{\text c}_3)=\frac{q-1}{q^2-4q+1}$ \\
\hline
		${\text m}_1\neq {\text c}_1$,${\text m}_2= {\text c}_2$,${\text m}_3\neq {\text c}_3$ & $\mu({\text m}_2)=\frac{1}{q-3}$ & $\nu ({\text m}_3,{\text c}_3)=\frac{q}{q^2-4q+1}$ \\
\hline
		${\text m}_1\neq {\text c}_1$,${\text m}_2\neq {\text c}_2$,${\text m}_3= {\text c}_3$ & $\mu({\text m}_2)=\frac{1}{q-3}$ & $\nu ({\text m}_3,{\text c}_3)=\frac{q-1}{q^2-5q+1}$ \\
\hline
		${\text m}_1\neq {\text c}_1$,${\text m}_2\neq {\text c}_2$,${\text m}_3\neq {\text c}_3$ & $\mu({\text m}_2)=\frac{1}{q-3}$ & $\nu ({\text m}_3,{\text c}_3)=\frac{q}{q^2-5q+1}$ \\
\hline

	\end{tabular}
\end{center}

Also the last encryption step is approximately perfect for large $q$.

To show the completeness of the Minkowski cipher in the sense of definition~\ref{dfn:complete_cryptographic_scheme}, we will look at the special case of a number field $\mathbf{F}$ with $\charact\mathbf{F}=2$. The points ${\text p}\in \boldsymbol{H}^\infty$ can then be described in the form ${\text p}(x,y)$ and $x\in \Z_2^n$, $y\in \Z_2^n$. A message of $2n$ bits length can be understood as the point ${\text p}(x,y)$ with coordinates $x$ and $y$. The $i$-th bit of the message shall be the $i$-th position in the representation of the point ${\text p}(x,y)$, which would be the $i$-th component of the vector $x$ for $i=1,...,n$ and the $(i-n)$-th component of the vector $y$ for $i=n+1,...,2n$. 

Let ${\text m}$ be a message point and ${\text c}$ be the corresponding cipher text point when using the key point ${\text k}$. The encryption is complete in the sense of definition~\ref{dfn:complete_cryptographic_scheme}, if there is a representation of ${\text m}$ where a change in position $j$ of ${\text c}$ is caused by a change in position $i$ of ${\text m}$, $i,j\in \{1,2,...,2n\}$. 

Two cases have to be considered. The indices $i$ and $j$ may affect the same coordinate of both ${\text m}$ and ${\text c}$. Without restricting generality we assume the $x$-coordinate. Alternatively, the first index $i$ may affect one coordinate, say the $x$-coordinate of ${\text m}$, the second index $j$ may then affect the $y$-coordinate of ${\text c}$. 

We have a look at the second case first. Let ${\text m}(x,y)$ and ${\text c}(u,v)$ be the message and ciphertext points and their coordinates. Furthermore, let $e_i$, $i\in \{1,...,n\}$ be the unit vectors in $\Z_2^n$. We transform the index $j$ to $j\rightarrow j-n$. Then ${\text m}'(x+e_i,y)$ and ${\text c}'(u,v+e_j)$ are the message and ciphertext points with changes in positions $i$ and $j$. 

It has to be shown that for any key ${\text k}$ it is possible to find two points ${\text m},{\text c}$ as message and cipher text in a way that the encryption function $f\in F$ holds $f({\text k},{\text m})={\text c}$ and $f({\text k},{\text m}')={\text c}'$. Due to the definition of the encryption function in definition~\ref{dfn:minkowski_cipher} the cases $[{\text m}]\cap [{\text k}]\leq 2$, $[{\text m}]_1=[{\text k}]_1$ and $[{\text m}]_2=[{\text k}]_2$ need to be distinguished. Without restricting generality we will assume that ${\text k}=(0,0)$, otherwise the Minkowski plane can be transformed accordingly. 

\begin{itemize}
	\item [${[{\text m}]}\cap {[{\text k}]}\leq 2$] Due to the way, the keys of the Minkowski cipher have been chosen in definition~\ref{dfn:minkowski_cipher}, we have to find two lines ${\text K},{\text K}'$ in $\boldsymbol{G}_1^\infty\cup \boldsymbol{G}_2^\infty\cup \boldsymbol{X}^\infty$ over the field $\Z_2^n$ with ${\text m},{\text c},{\text k}\in {\text K}$ and ${\text m}',{\text c}',{\text k}\in {\text K}'$. 
	
	  Our assertion is equal to 
	
	\begin{equation}\label{assertion_completeness}
	\begin{split}
	\determ({\text m},{\text c})&=0 \\
	\determ({\text m}',{\text c}')&=0
	\end{split}
	\end{equation}
	
	Substituting the coordinates for ${\text m},{\text m}',{\text c},{\text c}'$ results in:
	
	\begin{equation}\label{assertion_completeness_with_coordinates}
	\begin{split}
	xv-uy&=0 \\
	xv+xe_j+ve_i+e_ie_j-uy&=0
	\end{split}
	\end{equation}
	
	Or:
	
	\begin{equation}
	\begin{split}
	xv-uy&=0 \\
	xe_j+ve_i+e_ie_j&=0
	\end{split}
	\end{equation}
	
	It is easy to provide coordinates for ${\text m}(x,y)$ and ${\text c}(u,v)$ that satisfy these equations.
	
	\item[${[{\text m}]}_1= {[{\text k}]}_1$] According to the construction of the encryption function in definition~\ref{dfn:minkowski_cipher} the cipher text points ${\text c},{\text c}'$ are both on the line $[{\text k}]_2$. So the conditions in equation~(\ref{assertion_completeness}) are met.
	
	\item[${[{\text m}]}_2= {[{\text k}]}_2$] It follows that ${\text c}$ is on $[{\text k}]_1$. So the first condition of equation~(\ref{assertion_completeness}) is met. To fulfill also the seconde one, we have to find a line ${\text K}'$ with ${\text m}',{\text c}',{\text k}\in {\text K}'$. Again values can be found for ${\text m}$ and ${\text c}$ so that following equation is fulfilled:
	
		\begin{equation}
	\determ({\text m}',{\text c}')=0
	\end{equation}
	
\end{itemize}

This concludes the examination of the case that the indices $i$ and $j$ affect different coordinates of ${\text m}$ and ${\text c}$. 

To show the first case we again assume that the two points ${\text m}(x,y)$ and ${\text c}(u,v)$ and their coordinates as given. Let now ${\text m}'(x+e_i,y)$ and ${\text c}'(u+e_j,v)$ be the altered points. Again the three different ways of encryption need to be distinguished:

\begin{itemize}
	\item [${[{\text m}]}\cap {[{\text k}]}\leq 2$] Once again it has to be shown that condition (\ref{assertion_completeness}) is fulfilled. In analogy to the procedure shown above we reach following conditions:

\begin{equation}
\begin{split}
xv-uy&=0 \\
xv+e_iv-yu-ye_j&=0
\end{split}
\end{equation}

Which can be simplified to:

\begin{equation}
\begin{split}
xv-uy&=0 \\
e_iv-ye_j&=0
\end{split}
\end{equation}

Again it is easy to find suitable points ${\text m}$ and ${\text c}$ to meet these conditions for all possible $i,j$.

	\item[${[{\text m}]}_1= {[{\text k}]}_1$] In this case the message has the coordinates ${\text m}(x,0)$. From definition~\ref{dfn:minkowski_cipher} of the Minkowski cipher follows that the cipher text has the coordinates ${\text c}(0,v)$. Hence the condition~(\ref{assertion_completeness}) leads to the following:
	
\begin{equation}\label{minkowski_n_complete}
\begin{split}
xv&=0 \\
xv+e_iv&=0
\end{split}
\end{equation}
	
	Since both $x$ and $v$ cannot be $0$ due to the requirements for the keys of the Minkowski cipher in definition~\ref{dfn:minkowski_cipher}, equation~(\ref*{minkowski_n_complete}) has no legitimate solution.

	\item[${[{\text m}]}_2= {[{\text k}]}_2$] In this case the message has the coordinates ${\text m}(0,y)$. From definition~\ref{dfn:minkowski_cipher} of the Minkowski cipher follows that the cipher text has the coordinates ${\text c}(u,0)$. Hence the condition~(\ref{assertion_completeness}) leads to the following:

\begin{equation}\label{minkowski_n_complete_1}
(0+e_i)0-y(u+e_j)=0
\end{equation}

Since $y\neq 0$ and $u\neq -e_j$, as ${\text c}'\neq {\text k}$ due to the requirements for the keys of the Minkowski cipher in definition~\ref{dfn:minkowski_cipher}, equation~(\ref*{minkowski_n_complete_1}) has no legitimate solution.

\end{itemize}

As a summary we showed that the Minkowski cipher with the encryption functions introduced in definition~\ref{dfn:minkowski_cipher} is not complete in accordance with definition~\ref{dfn:complete_cryptographic_scheme}.

\section{Resume and Outlook}
We introduced encryption functions on the geometric structures of the Laguerre and the Minkowski geometry and showed that the former fulfills both the requirements of  perfectness in the sense of Shannon~\cite{Sha49} and completeness according to~\cite{KD79}. The latter is also perfect in first approximation but doesn't have the property of compleness, at least when using the encryption functions as defined in this paper. Further research can be done in the Minkowski geometry to find improved encryption functions. One way would be to apply a similar approach to that one used in the Laguerre geometry, i.e. select generators as sets of possible messages and ciphertexts.

\end{document}